\documentclass[12pt]{article}
\usepackage{graphicx}
\begin{document}
\title{Effective Compton Cross Section in Non-Degenerate High Temperature Media}
\author{F. Shekh-Momeni\footnote{e-mail: fmomeni@mehr.sharif.edu}
and J. Samimi \footnote{e-mail: samimi@sharif.edu}\\ \\
Department of Physics, Sharif University of Technology,\\ Tehran,
P.O.Box: 11365-9161, Iran} \maketitle
 \begin{abstract}
 The effective compton cross section in a non-degenerate
 plasma($n\ll\{\frac{(kT/c)^{2}+2mkT}{h^{2}}\}^{^{3/2}}$)
 is investigated in a wide range of temperatures. The results show
 a decreasing behavior with temperature especially for $kT\gg m_{e}c^{2}$. The results
may be important in phenomena like accretion discs or
ultra-relativistic blast waves in GRB models, where the emitted
radiation has to pass through a medium containing high energy
electrons.\\

 {\it keywords}:Compton scattering, high energy physics, gamma-ray
bursts, accretion
\end{abstract}
\section{Introduction}

Compton scattering is one of the most important phenomena in
astrophysics that affects the received flux from terrestrial
objects. In a non-degenerate low temperature plasma in which
$\frac{n^{2/3}h^{2}}{2\pi\:m_{e}}\ll kT\ll m_{e}c^{2}$
\cite{Phillips}, the free electrons can be considered effectively
at rest, provided that the photon energy $E=h\nu$ to be much
greater than $kT$. In such a situation the Klein-Nishina formula
for Compton scattering accurately works. The Klein-Nishina formula
is only dependent on the energy of the incoming photon as measured
in the rest frame of the electron \cite{Stecker}, \cite{Harwit} :
\begin{equation}\label{eq:a2}
\sigma_{_{c}}=\sigma_{_{T}}.\frac{3}{4}
 \{\frac{1+\epsilon}{\epsilon^{2}}[\frac{2(1+\epsilon)}{1+2\epsilon}
-\frac{1}{\epsilon}\ln(1+2\epsilon)]+\frac{1}{2\epsilon}\ln(1+2\epsilon)-
\frac{1+3\epsilon}{(1+2\epsilon)^{^{2}}}
\end{equation}
in which $\epsilon$ is the ratio of the photon energy $E$ to the
electron rest energy $m_{e}c^{2}$ (Fig.(1)). But at high
temperatures where the electron mean energy is greater than or
comparable to $m_{e}c^{2}$, the energy of the photon is highly
different relative to different individual electrons. So, since
the total Compton cross section is a function of the photon
energy in the rest frame of the electrons, one would expect an
effective cross section due to the thermal distribution of
electrons. The aim of this paper is to obtain an effective cross
section for Compton scattering in a wide range of plasma
temperatures and photon energies. In Sec. 2 the averaging method
is introduced. In Sec. 3 the momentum distribution of electrons in
a wide range of temperatures is discussed and the normalization
constant of the distribution is evaluated as a function of plasma
temperature, and then, a numerically computable form for the
effective cross section of Compton scattering is obtained. The
results are presented in Sec. 4, as well as a discussion on their
importance in GRBs.

\section{Averaging formalism}

Consider a monochromatic photon beam with energy $E=h\nu$ passing
through a plasma that its temperature $T$ and its electron number
density $n_{e}$ satisfy the non-degeneracy condition
$n\ll\{\frac{(kT/c)^{2}+2mkT}{h^{2}}\}^{^{3/2}}$ . Now, we define
the effective cross section in Compton scattering
$\sigma_{_{C,eff}}$ as:
\begin{equation}\label{eq:a2}
\frac{dN}{N}=\sigma_{_{C,eff}}.n_{e}.dl
\end{equation}
where $\frac{dN}{N}$ denotes the fraction of photons scattered in
a distance $dl$. Now, let's consider the fraction of electrons
$\frac{\delta n_{e}}{n_{e}}$ that move in the solid angle
$d\Omega$ and have a momentum magnitude between p and p+dp. The
fraction has the value:
\begin{equation}\label{eq:a2}
\frac{\delta
n_{e}}{n_{e}}=\frac{d^{3}f}{d\overrightarrow{p}^{3
}}\:p^{2}\:dp\:d\Omega
\end{equation}
where $\frac{d^{3}f}{d\overrightarrow{p}^{3}}$ is the probability
density for an electron to have momentum $\overrightarrow{p}$.
Let's give attention to the photons that scatter by these
electrons. Denoting them by $\delta N$, we can write :
\begin{equation}\label{eq:a2}
\frac{dN}{N}=\frac{\int\delta N}{N}
\end{equation}
which means that the total scattering is the result of integration
over all possible scatterings by the electrons having different
momenta. To find $\frac{\delta N}{N}$, we make a relativistic
transformation to the comoving frame of the electrons  denoted by
$\delta n_{e}$ (Eqn.(3)), hereafter called as "the prime frame".
The scattering of the photons by these electrons is an "event",
as it mean in the theory of relativity. So, the quantity
$\frac{\delta N}{N}$ is an invariant and once evaluated in a
specific frame, its magnitude can be used in all other frames.
Since the electrons denoted by $\delta n_{e}$ are at rest in
prime frame (which is their own comoving frame) we can use
Klein-Nishina cross section (Eqn. (1)) to evaluate $\frac{\delta
N}{N}$ in this frame:
\begin{equation}\label{eq:a2}
\frac{\delta N}{N}=\sigma_{c}(\epsilon^{\prime}).\delta
n_{e}^{\prime}.dl^{\prime}
\end{equation}
where, $\epsilon^{\prime},\:\delta n_{e}^{\prime}\:,$ and
$dl^{\prime}$ correspond the quantities $\epsilon,\delta n_{e}$
and $dl$ in the the prime frame, so that :
\begin{equation}\label{eq:a2}
\epsilon^{\prime}=\epsilon\:\gamma\:(1-\beta\:\cos\theta)
\end{equation}
\begin{equation}\label{eq:a2}
\delta n_{e}^{\prime}=\frac{\delta n_{e}}{\gamma}
\end{equation}
\begin{equation}\label{eq:a2}
dl^{\prime}=dl\:\gamma\:(1-\beta\:\cos\theta)
\end{equation}
where $\gamma$ and $\beta$ are the Lorentz factor and the speed of
the mentioned electrons respectively and $\theta$, the polar angle
that the electrons make with the photon beam direction, all
measured in laboratory frame. Using Eqs.(6) to (8), Eqn.(5) can
be written as:
\begin{equation}\label{eq:a2}
\frac{\delta
N}{N}=\sigma_{c}(\epsilon\:\gamma\:(1-\beta\:\cos\theta)).\delta
n_{e}.(1-\beta\:\cos\theta)\:dl
\end{equation}
and considering Eqn.(4), we see:
\begin{equation}\label{eq:a2}
\frac{dN}{N}=[\int\int\sigma_{c}(\epsilon\:\gamma\:(1-\beta\:\cos\theta))
.(1-\beta\:\cos\theta)
.\frac{d^{3}f}{d\overrightarrow{p}^{3
}}\:p^{2}\:dp\:d\Omega].n_{e}.dl
\end{equation}
where we used Eqs.(3) and (9). Comparing Eqn.(10) with Eqn.(2) we
obtain finally:
\begin{equation}\label{eq:a2}
\sigma_{_{C,eff}}=\int\int\sigma_{c}(\epsilon\:\gamma\:(1-\beta\:\cos\theta))
.(1-\beta\:\cos\theta)
.\frac{d^{3}f}{d\overrightarrow{p}^{3 }}\:p^{2}\:dp\:d\Omega
\end{equation}

To use this expression in evaluating $\sigma_{_{C,eff}}$, firstly
we need to find $\frac{d^{3}f}{d\overrightarrow{p}^{3 }}$
explicitly. The next section is devoted to this task.

\section{Normalization Constants}
The probability density for an electron to have a momentum
$\overrightarrow{p}$ is proportional to the Boltzman factor
$e^{-E/kT}$, so that:
\begin{equation}\label{eq:a2}
\frac{d^{3}f}{d\overrightarrow{p}^{3}}=C(T)\:e^{-E(p)/kT}
\end{equation}
We want to find the normalization constant C(T) which has the
following form:
\begin{equation}\label{eq:a2}
C(t)=\{\:\int_{0}^{\infty}e^{-E(p)/kT}\:4\pi\:p^{2}\:dp\:\}^{-1}
\end{equation}
Assuming the electrons to be free we simply have:
\begin{equation}\label{eq:a2}
E(p)=\sqrt{p^{2}c^{2}+m_{e}^{2}c^{4}}
\end{equation}
It is well known that the integration has simple analytic
solutions at two extreme temperature limits namely,
non-relativistic ($kT\ll m_{e}c^{2}$) and ultra-relativistic
($kT\gg m_{e}c^{2}$) temperatures:
\begin{equation}\label{eq:a2}
C(T)=\left\{\begin{array}{rcl}
   (2\pi\:m_{e}\:kT)^{-3/2}\:\:\:\:\:\:\:for\:\:\:\:\:kT\ll m_{e}c^{2}\\
   \frac{1}{\pi}\:(\frac{kT}{c})^{-3}\:\:\:\:\:\:\:for\:\:\:\:\:kT\gg m_{e}c^{2}
    \end{array}\right.\\
\end{equation}
But at intermediate temperatures there is no analytic solution for
the integration. So it must be solved numerically. The obtained
numerical results are shown in Fig.(3). It can be seen that C(t)
has a wide range of orders of magnitude. Really in practice we
had to take exact care in parameterizing the quantities, otherwise
the numerical computation would have been impossible. The
procedure used in parameterizing the quantities maybe worth
mentioning. the range of the integration in Eqn.(13) is from zero
to infinity. Therefore, we must concern ourselves with the
momentum $p_{max}$ in which the integrand $p^{2}\:e^{-E(p)/kT}$
reaches its maximum value. This is described by the equation:
\begin{equation}\label{eq:a2}
\frac{d}{dp}\:(p^{2}\:e^{-E(p)/kT})\mid_{p=p_{m}}=0
\end{equation}
which has the solution :
\begin{equation}\label{eq:a2}
p_{max}(T)=\sqrt{2}\:\{1+\sqrt{1+(\frac{m_{e}c^{2}}{kT})^{2}}\:\:\}\:\:\frac{kT}{c}
\end{equation}
Now, we define a number of non-dimensional parameters as the
following :
\begin{equation}\label{eq:a2}
\xi\equiv\frac{p}{p_{max}}
\end{equation}
\begin{equation}\label{eq:a2}
\tau\equiv\frac{kT}{m_{e}c^{2}}
\end{equation}
\begin{equation}\label{eq:a2}
c(\tau)\equiv C(T)\:\:p_{max}^{3}(T)
\end{equation}
\begin{equation}\label{eq:a2}
g(\tau)\equiv\sqrt{2}\:\tau\:\{1+\sqrt{1+\frac{1}{\tau^{2}}}\:\:\}
\end{equation}
\begin{equation}\label{eq:a2}
\alpha(\xi,\tau)\equiv
\frac{E(p)}{kT}=\frac{1}{\tau}\{\:\sqrt{1+\xi^{2}\:g(\tau)^{2}\:\:}\:\:-1\:\}
\end{equation}
and use them to convert Eqs.(12) and (13) to non-dimensional
forms below :
\begin{equation}\label{eq:a2}
\frac{d^{3}f}{d\overrightarrow{\xi}^{3
}}=c(\tau)\:e^{-\alpha(\xi,\tau)}
\end{equation}
\begin{equation}\label{eq:a2}
c(\tau)=\{\int_{0}^{\infty}\:4\pi\:\xi^{2}\:e^{-\alpha(\xi,\tau)}\:d\xi\}^{-1}
\end{equation}
Considering Eqn.(12), and Eqs.(17) to (23), it can be seen that
the quantity:
\begin{equation}\label{eq:a2}
\frac{df}{d\xi}\equiv
4\pi\:\xi^{2}\:\frac{d^{3}f}{d\overrightarrow{\xi}^{3
}}=4\pi\:c(\tau)\:\xi^{2}\:e^{-\alpha(\xi,\tau)}
\end{equation}
is naturally normalized:
\begin{equation}\label{eq:a2}
\int\frac{df}{d\xi}\:d\xi\:=1
\end{equation}
and takes its maximum value at $\xi=1$ which is independent of
$\tau$. Really, the parameterizing procedure was designed so that
$\frac{df}{d\xi}$ reach to its maximum at a point independent of
$\tau$, otherwise the wide range of $\tau$ values would make the
numerical computations impossible or at least inconvenient.

The results are shown in Fig.(2), and as expected, the
non-dimensional normalization constant $c(\tau)$ is a quantity of
order of one. Having these results in hands and using  Eqs.(17),
(19) and (20), the actual normalization constant $C(T)$ can be
obtained. the results are shown in Fig.(3).

Now, all is prepared to return to Eqn.(11) and rewrite it after
some algebra in terms of the defined parameters:
\begin{equation}\label{eq:a2}
\frac{\sigma_{c,eff}}{\sigma_{T}}=2\pi
c(\tau)\int_{\theta=0}^{\pi}\int_{0}^{\infty}
\frac{1}{\sigma_{T}}\sigma_{c}(\epsilon\gamma(1-\beta\cos\theta)).(1-\beta\cos\theta)
\xi^{2}e^{-\alpha(\xi,\tau)}\:\sin\theta\:d\theta\:d\xi
\end{equation}
Noting that $\gamma$ and $\beta$ in this relation must be
considered as functions of $\xi$ and $\tau$. So, let's write them
firstly in terms of momentum p:
\begin{equation}\label{eq:a2}
\gamma=\sqrt{(\frac{p}{m_{e}c})^{2}+1}
\end{equation}
\begin{equation}\label{eq:a2}
\beta=\frac{p\:c}{\sqrt{p^{2}c^{2}+m_{e}^{2}c^{4}}}
\end{equation}
and use Eqs.(17) to (19) and Eqn.(21) to obtain:
\begin{equation}\label{eq:a2}
\gamma=\sqrt{1+\xi^{2}g(\tau)^{2}}
\end{equation}
\begin{equation}\label{eq:a2}
\beta=\frac{\xi\:g(\tau)}{\sqrt{1+\xi^{2}g(\tau)^{2}}}
\end{equation}
Now the numerical evaluation of Eqn.(27) is straight forward. The
obtained results are presented and discussed in the next section.

\section{Results and Discussion}

As it can be seen in the three-dimensional plot of Fig.(4), and
in the contour plot of Fig.(5), the effective cross section is
closely the same as the $\sigma_{c}$ (Fig.(1)) at temperatures
$kT<m_{e}c^{2}$. This is an anticipated result. The deviation
begins when $kT>m_{e}c^{2}$, with an overall shift to left at
higher and higher temperatures. This behavior could have been
expected too. The Compton cross section decreases by increasing of
photon energy (Fig.(1)), so, for a photon with energy $E=h\nu$,
the more the temperature is, the more energy in rest frame of
individual electrons it has, and furthermore, the less the
effective cross section is expected. This simple reasoning may
seem wrong of course. Because though it is correct for the
electrons moving toward the photon beam but in the rest frame of
the electrons moving along the beam the energy of a photon would
be less than the corresponding value as measured in laboratory
frame and it decreases more and more in higher and higher
temperatures, and so, in contrary one may expect the effective
cross section to be less and less, as long as these electrons are
concerned. Really, the final answer can be found in Eqs.(5) and
(8).For ultra-relativistic electrons the distance $dl^{\prime}$
as seen by the electrons moving in the same direction as the beam
one, is $4\gamma^{2}$ times less than the corresponding value for
ones moving in the opposite direction. So, as can be seen in Eqn.
(5), despite the greater cross section they have individually,
their share in scattering happens to be less than that other
electrons, so that in high temperatures the effective cross
section as a whole is determined by the electrons moving toward
the photon beam. In Fig.(6) the portion of these two groups of
electrons in Compton scattering is compared with each other.

In a GRB model prepared by authors \cite{grb}, a collimated
ultra-relativistic ejecta with a Lorentz factor $\Gamma \sim 1000$
collides with a dense cloud surrounding the stellar engine. The
photons emitted from the shocked medium has to pass through the
cloud before entering free space, but the opacity of the cloud is
extremely too high to let them escape. Really, in the model the
cloud thickness $L$, and its density $n$, are of order $10^{13}
cm$ and $10^{17} cm^{-3}$ respectively, which result in an opacity
high to $\sigma_{_{T}} n L \sim 10^{6}$ which obviously prevents
any radiation to escape. But, since in external shock models for
GRBs (see \cite{piran} and \cite{Katz} for a review) the particles
in the shocked medium are expected to have mean energies of order
$\frac{u}{4\:\Gamma\:n}=\Gamma\:m_{p}\:c^{2}\sim10^{6}\:m_{e}\:c^{2}$
, the real optical depth of the medium for $Mev$ photons would be
$10^{-6}$ times less than what might be roughly expected in
beginning (Table (1) ), and consequently a photon radiated from
the shocked matter may succeed to go out of it without being
scattered, provided that its passage make an angle larger than
$(\sqrt{\frac{5}{3}}\Gamma)^{-1}$ with the velocity vector of
shocked matter so that it remain in the shocked medium by the time
it cross the opaque cloud. So, we concluded that if the shock
front succeed to cross the cloud it might be seen and so might
really make a GRB, and if it do not and stop in the cloud, its
photons would not be able to cross the cool dense cloud and the
phenomenon must be considered a $\textit{failed}$ GRB. Though the
presented reasoning may seem to be restricted only to a medium in
thermal equilibrium but it must be correct for all media in which
the mean energy of electrons is of order of
$10^{6}\:m_{e}\:c^{2}$. for example in GRB models, the electrons
in the shocked matter are assumed to have a power law
distribution:
\begin{equation}\label{eq:a2}
N(\gamma_{e})\propto\gamma_{e}^{2}\:\:\:\:\:\:\:\:\:\:\:\:\:for
:\:\:\:\:\:\:\:\:\:\:\:\gamma_{e}>\gamma_{e,min}
\end{equation}
By repeating the averaging procedure introduced in this paper the
effective Compton cross section for such a distribution can be
found \cite{momeni2}. The obtained results are comparable in
orders of magnitude to ones presented in Table (1) and shown in
Figs.(4) and (5), replacing $\tau$ with mean Lorentz factor of
electrons.

\begin{figure}
   \centering
   \includegraphics[width=8cm]{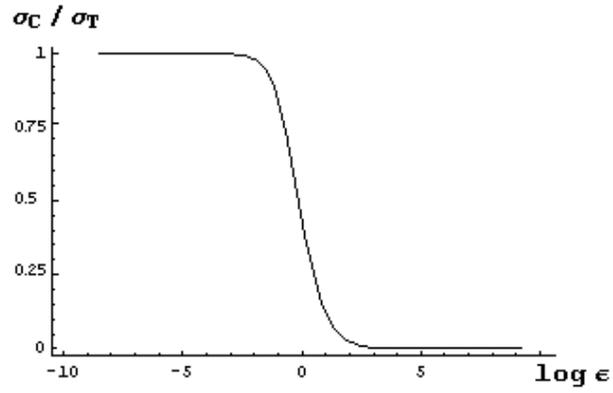}
      \caption{the cross section in Compton scattering evaluated using Eqn.(1) }
       \label{fit}
   \end{figure}

   \begin{figure}
   \centering
   \includegraphics[width=8cm]{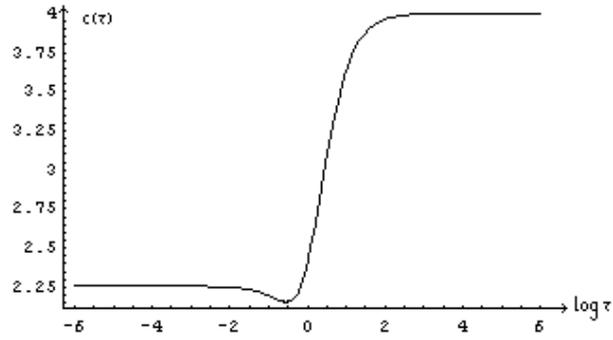}
      \caption{The parameterized non-dimensional normalization constant $c(\tau)$ verses $\log \tau$}
       \label{fit}
   \end{figure}

\begin{figure}
   \centering
   \includegraphics[width=8cm]{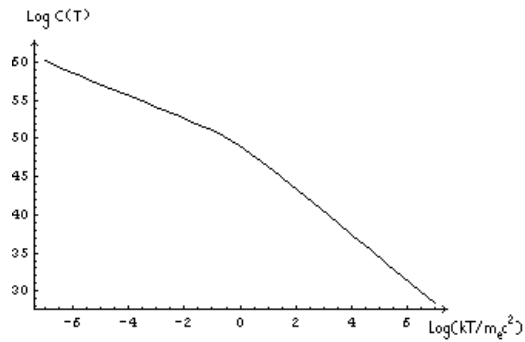}
      \caption{The normalization constant $C(T)$ verses temperature}
       \label{fit}
   \end{figure}

   \begin{figure}
   \centering
   \includegraphics[width=10cm]{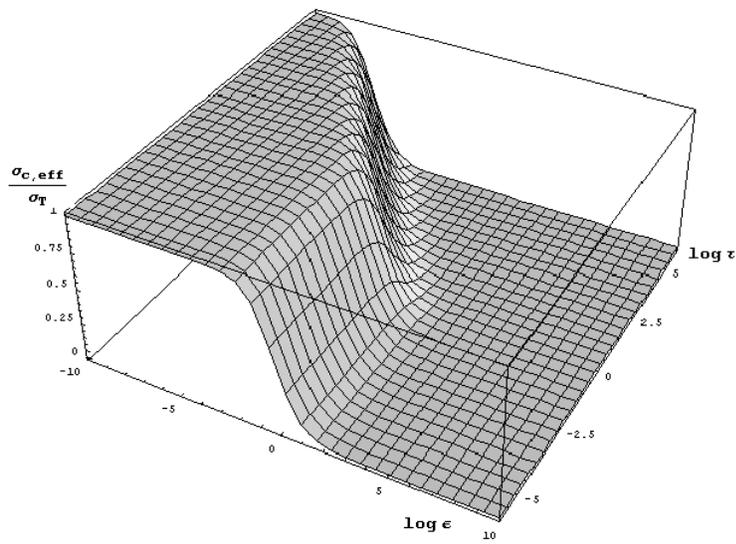}
      \caption{The effective Compton scattering cross section
       (normalized by Thomson cross section) as a function of
        $log (\epsilon\equiv \frac{E}{m_{e}c^{2}})$ and
        $log (\tau \equiv \frac{kT}{m_{e}c^{2}})$, where $E$ and $T$ are the photon energy and
        the temperature of non-degenerate gas of free electrons (see the text)}
       \label{fit}
   \end{figure}

   \begin{figure}
   \centering
   \includegraphics[width=12cm]{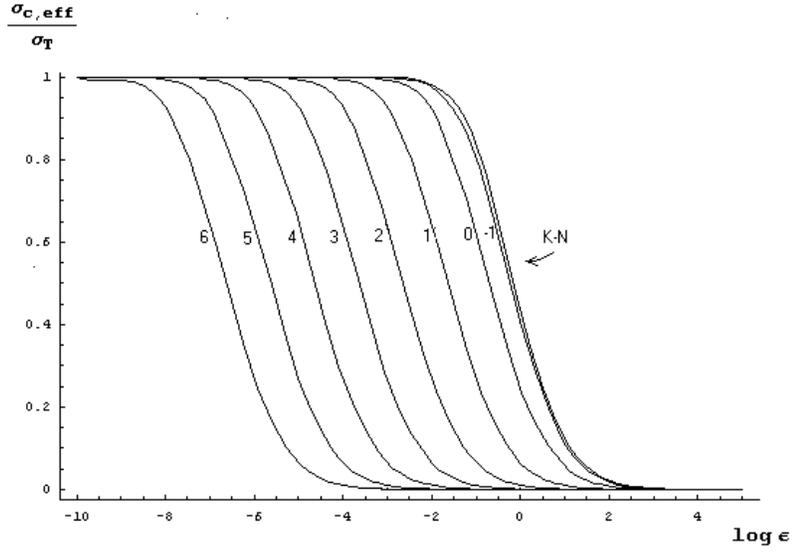}
      \caption{The effective Compton scattering cross section
      verses  $log (\epsilon\equiv \frac{E}{m_{e}c^{2}})$
       (normalized by Tompson cross section). the label on each curve is the logarithm of
       $\frac{kT}{m_{e}c^{2}}$. In low temperatures $kT \ll m_{e}c^{2}$ the curve
       approach to the most right curve
       corresponding to the traditional Compton cross section as evaluated using
        Klain-Neshina formula (Fig.(1)). In higher an higher temperatures
        the overall behaviour is a shift toward left. (see the
        text)}
       \label{fit}
\end{figure}

\begin{figure}
   \centering
   \includegraphics[width=12cm]{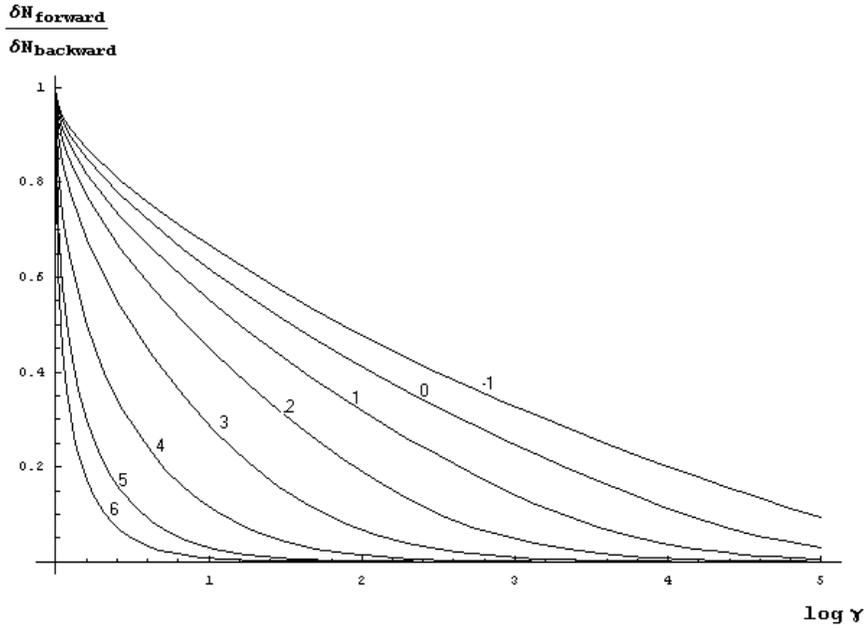}
      \caption{The ratio of photons scattered by the electrons moving in the same
      direction as the beam to ones moving in the opposite direction. The horizontal axis is
      the logarithm of electron Lorentz factor, and the number near each
       curve is the logarithm of
       $\epsilon\equiv\frac{E}{m_{e}c^{2}}$ ,with $E$ being the incoming photon energy.
       It can be seen that the ratio equals one when electrons move slowly, while it
       approaches to zero for ultra-relativistic electrons, showing that in
       ultra-relativistic temperatures the electrons moving in the
       opposite direction have the most share in Compton
       scattering (Sec.4).}
       \label{fit}
   \end{figure}

   \newpage

   \begin{figure}
   \begin{picture}(200,200)(0,0)
   \includegraphics{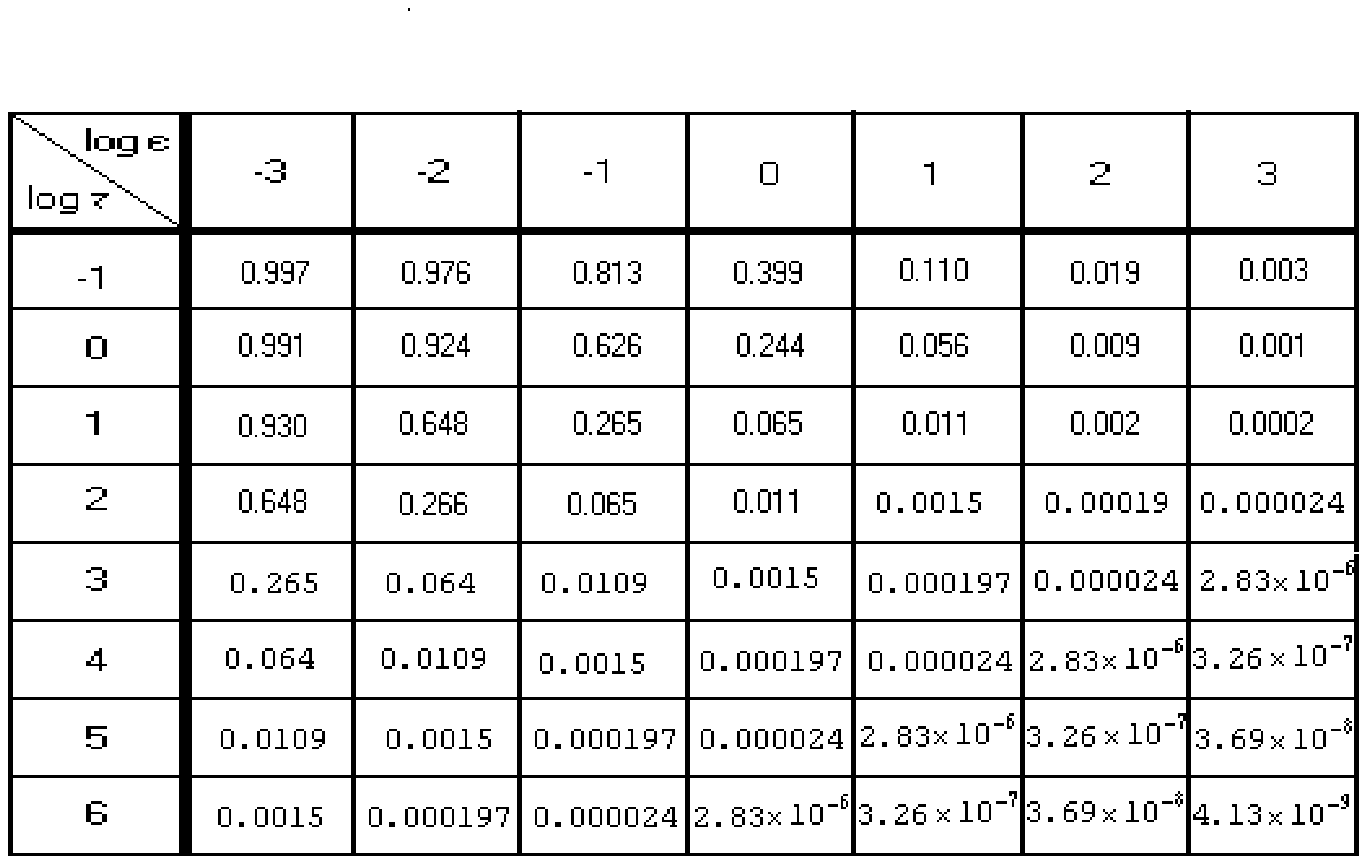}

  \end{picture}
     \caption{\textbf{TABLE I} The effective Compton cross section }
       \label{fit}
   \end{figure}

 \end{document}